# Foundations of the theory of gravity with a constraint. Gravitational energy of macroscopic bodies

Alexander P. Sobolev[1,§] and Aleksey Sobolev[2,‡,*]

[1]ORCID 0000-0003-0649-7305
[2]ORCID 0009-0000-6304-7883
[§]Moscow Institute of Physics and Technology, Russian Federation
[‡]Independent researcher
[*]Author to whom correspondence should be addressed: aleksey.sobolev01@gmail.com

## Abstract

The paper considers a set of equations describing the static isotropic gravity field of a macroscopic body within the framework of the theory of gravity with a constraint. A general approximate solution of these equations is obtained. The solution exists only at certain values for the three integration constants. The out-of-body metric coincides with the Schwarzschild metric, but, unlike the general relativity theory (GR), the curvature tensor invariants have a certain finite value everywhere.

## 1. Introduction

In 2022, the article 'Foundations of a Theory of Gravity with a Constraint and Its Canonical Quantization' was published [1] (a corrected version of this paper [2]). The distinguishing feature of this theory, compared to Einstein-Hilbert's general relativity theory (GR), is the introduction of a constraint among the metric tensor components, replacing the general covariance hypothesis. The inclusion of this constraint enabled the determination of the symmetric energy-momentum tensor of the gravity field, along with its entropy and temperature. Consequently, the gravity field, in this scenario, exhibits properties characteristic of material media and serves as the primary source of the entire energy of the Universe. It has been demonstrated in [1] that the cosmological principle (the hypothesis of homogeneity and isotropy of space) is rigorously satisfied in such a Universe prior to the emergence of matter.

At the moment of the onset of the Universe's evolution, the energy density of the gravity field was equal to zero, and its tension was negative [1]. This served as a trigger for the accelerated increase in the initially minimal volume factor and the energy density of the field itself. The energy density, within the time of about $10^{-12}$ seconds, reached a value of $\sim 10^{50}$ J·m$^{-3}$ [1], which made it possible to launch a process of inception of matter fields due to the gravity field energy. Thus, the centuries-old problem of the creation of matter from nothing (creation ex nihilo) was resolved.

Building upon this foundation, a unified evolution model was developed for a non-stationary globally homogeneous large-scale structure encompassing the modern, early, and very early Universe [2]. Within this model, it was demonstrated that the current global energy density in the Universe primarily comprises 94.5% of the gravity field's energy density, with only 5.5% attributed to all the known types of matter. Notably, the model's calculated results align with observable astronomical data, eliminating the need for hypotheses such as 'dark energy', 'dark matter', and 'inflatons'.

In addition, a wave equation for the gravity field was formulated, and a nonstationary wave function for the very early Universe was constructed [2]. It is shown that the origination of the Universe was random in nature and that the gravity field has only a continuous energy spectrum, while the spectrum of discrete energy levels is absent.

Hence, the theory of gravity with a constraint [2] provides solutions to several long-standing problems that remain unresolved within the framework of GR. The predictive capabilities of GR are confined to the limits of our solar system. To explain observation results concerning intragalactic and extragalactic objects, which deviate from GR predictions, one must hypothesize the existence of undiscovered forms of matter in the Universe (forms not observed within our solar system). While experimentally validated within the solar system, GR's divergence from observed data in other regions suggests the need for alternative gravity theories. Regardless of conceptual disparities, any new theory of gravity should align with GR's results to compare with experimental data. Therefore, considering the application of the theory of gravity with a constraint for a static non-homogeneous metric becomes necessary for comparison with experimental data.

Proceeding from Hilbert's axioms [3] (in the presently accepted designations), the gravitational action in the theory of gravity with a constraint was initially assigned as follows [1]

$$S_{gr} = -\frac{c^3}{16\pi G}\int (R+Q)\sqrt{-g}d^4x, \quad Q = \frac{1}{\sqrt{-g}}\frac{\partial\sqrt{-g}}{\partial x^\mu}g^{\mu\nu}\frac{\partial\Phi}{\partial x^\nu} \quad (1)$$

where $R$ is the scalar curvature, $\Phi(x^\mu)$ is the Lagrange multiplier.

The expression $Q$ is a restrictedly covariant scalar because it contains a partial derivative of the volume factor instead of the covariant derivative. Therefore, unlike the Hilbert action, integration is defined not on manifolds but only on manifolds with an edge. The presence of an additional term in the action, in addition to the scalar curvature, yields its contribution at the variation. The variation of the action (1) with respect to the Lagrange multiplier yields the constraint equation.

$$\frac{\partial}{\partial x^\mu}\left(g^{\mu\nu}\frac{\partial\sqrt{-g}}{\partial x^\nu}\right) = 0$$

The variation with respect to the metric leads to the appearance the energy-momentum density tensor of the gravity field $(\varepsilon_{gr})_{\mu\nu}$ in the Einstein-Hilbert equations, along with the energy-momentum density tensor of the matter $(\varepsilon_{mat})_{\mu\nu}$.

$$R_{\mu\nu} - \frac{1}{2}g_{\mu\nu}R = \frac{8\pi G}{c^4}(\varepsilon_{gr})_{\mu\nu} + \frac{8\pi G}{c^4}(\varepsilon_{mat})_{\mu\nu},$$

$$\frac{8\pi G}{c^4}(\varepsilon_{gr})_{\mu\nu} = -\frac{1}{2}\left[g_{\mu\nu}\frac{\partial}{\partial x^\rho}\left(g^{\rho\lambda}\frac{\partial\Phi}{\partial x^\lambda}\right) + \frac{1}{\sqrt{-g}}\frac{\partial\sqrt{-g}}{\partial x^\mu}\frac{\partial\Phi}{\partial x^\nu} + \frac{1}{\sqrt{-g}}\frac{\partial\sqrt{-g}}{\partial x^\nu}\frac{\partial\Phi}{\partial x^\mu}\right].$$

Further it was shown [1] that, unlike GR, the action (1) may attain its minimum only when the negative sign within the integral changes to a positive sign. In accordance with this, it is necessary to change the signs in the motion equations. Thus, the equations of the theory of gravity with a constraint, in the presence of matter, have the following form:

$$\frac{\partial}{\partial x^\mu}\left(g^{\mu\nu}\frac{\partial\sqrt{-g}}{\partial x^\nu}\right) = 0, \quad (2)$$

$$-R_{\mu\nu} + \frac{1}{2}g_{\mu\nu}R = -\frac{8\pi G}{c^4}(\varepsilon_{gr})_{\mu\nu} + \frac{8\pi G}{c^4}(\varepsilon_{mat})_{\mu\nu}, \quad (3)$$

$$\frac{8\pi G}{c^4}(\varepsilon_{gr})_{\mu\nu} = -\frac{1}{2}\left[g_{\mu\nu}\frac{\partial}{\partial x^\rho}\left(g^{\rho\lambda}\frac{\partial\Phi}{\partial x^\lambda}\right) + \frac{1}{\sqrt{-g}}\frac{\partial\sqrt{-g}}{\partial x^\mu}\frac{\partial\Phi}{\partial x^\nu} + \frac{1}{\sqrt{-g}}\frac{\partial\sqrt{-g}}{\partial x^\nu}\frac{\partial\Phi}{\partial x^\mu}\right]. \quad (4)$$

The gravity field will actually possess all the properties of the material medium only if two conditions are met [1]:

$$\Phi(x) \neq const, \quad g(x) \neq const. \quad (5)$$

In GR, the first of these conditions is not met, whereas in the unimodular theory of gravity, the second one is not met. Subsequently, the conditions of (5) will be deemed as observed.

## 2. Preamble

*The presence of material bodies (planets) violates the homogeneity of space and leads to changes in the curvature and energy density of the gravity field both inside and outside the bodies.*
Lemma. *For any static solid body, the energy of the gravity field created by it is equal in magnitude to its rest energy.*

Now we pass to compounded indices in equations (3)

$$R_\mu^\nu - \frac{1}{2}\delta_\mu^\nu R = \frac{8\pi G}{c^4}\left[(\varepsilon_{gr})_\mu^\nu - (\varepsilon_{mat})_\mu^\nu\right].$$

It's important to note that in this equation, the energy-momentum density appears with a minus sign, unlike in GR. According to the convoluted Bianchi identity [5, p.146], the covariant derivative of the right-hand side of this equation must be zero.

$$\frac{1}{\sqrt{-g}}\frac{\partial}{\partial x^\nu}\left(\sqrt{-g}\left[(\varepsilon_{gr})^\nu_\mu - (\varepsilon_{mat})^\nu_\mu\right]\right) - \frac{1}{2}\frac{\partial g_{\lambda\rho}}{\partial x^\mu}\left[(\varepsilon_{gr})^{\lambda\rho} - (\varepsilon_{mat})^{\lambda\rho}\right] = 0 \;.$$

If the index $\mu$ is assigned a value equal to zero, then, by virtue of the metric staticity, the last term of the equality turns to zero, and in this case

$$\frac{1}{\sqrt{-g}}\frac{\partial}{\partial x^\nu}\left(\sqrt{-g}\left[(\varepsilon_{gr})^\nu_0 - (\varepsilon_{mat})^\nu_0\right]\right) = 0 \;.$$

After integrating this ratio by 4-volume and transforming it into a hypersurface integral [4, ch. XI], we derive

$$\int \left[(\varepsilon_{gr})^\nu_0 - (\varepsilon_{mat})^\nu_0\right]\sqrt{-g}\,dS_\nu = const\;. \quad (6)$$

In the case of a static field, it follows from (4)

$$(\varepsilon_{gr})^\lambda_0 = -\frac{c^4}{16\pi G}\delta^\lambda_0 \frac{\partial}{\partial x^\mu}\left(g^{\mu\nu}\frac{\partial \Phi}{\partial x^\nu}\right).$$

For a static solid body with density $\varepsilon(x)$ [4, ch. IV, § 35]

$$(\varepsilon_{mat})^\lambda_0 = \varepsilon(x)u_0 u^\lambda\;,$$

where $u^\lambda = (1, 0, 0, 0,)$. Substituting these expressions into (6), we derive

$$\int \left[(\varepsilon_{gr})^0_0 - (\varepsilon_{mat})^0_0\right]\sqrt{-g}\,dS_0 = \int \left[(\varepsilon_{gr})^0_0 - (\varepsilon_{mat})^0_0\right]\sqrt{-g}\,d^3x = const\;.$$

Since the integral over the volume of the whole space covers the difference in energy densities, this equation can be considered as a definition of the gravitational energy of a resting solid body

$$E_{gr} = E_{mat} + const.$$

Considering that the gravity field becomes homogeneous in the absence of a material body, and its energy in this case is also equal to zero, one concludes that $const = 0$. Hence,

$$E_{gr} = -\frac{c^4}{16\pi G}\int \frac{\partial}{\partial x^\mu}\left(g^{\mu\nu}\frac{\partial \Phi}{\partial x^\nu}\right)\sqrt{-g}\,d^3x = E_{mat}\;. \quad (7)$$

### 3. Problem statement

The shortest way to derive the gravity field equations in this case is specifying an action and applying the principle of minimum.

Following the approach set forth in the section "Classic tests of Einstein's theory" [5, pp. 175–209], one should consider a static spherically symmetric metric (preserving the notations for the metric and coordinates adopted in the cited paper). In GR, the metric independent of coordinate $x^0$, using the coordinates transformation [4, ch. X]

$$x'^0 = x^0 + \phi(x^m)\;,\; x'^m = x^m\; (m = 1, 2, 3),$$

can be reduced to a static form with the components $g_{0m}$ equal to zero. Since this transformation is unimodular, a similar assumption is also true for the theory of gravity with a constraint.

By analogy with [5, p. 176], the most general expression for the space-time interval can be reduced to the form

$$ds^2 = F(r)(dx^0)^2 - \frac{G(r)}{r^2}(x^m \delta_{mn}\,dx^n)^2 - C(r)(dx^m \delta_{mn} dx^n),\; r = (x^m \delta_{mn}\,x^n)^{1/2}\;. \quad (8)$$

(Latin indices take values 1, 2, 3). In contrast to [5], Kronecker symbols $\delta_{mn}$ are used here for notation. Thus, in this case, the metric tensor $g_{\mu\nu}$ (Greek indices take the values 0, 1, 2, 3) has the following form

$$g_{00} = F(r),\quad g_{0m} = 0,\quad g_{mn} = -C(r)\delta_{mn} - G(r)\delta_{mk}\delta_{nl}\frac{x^k x^l}{r^2},\quad (9)$$
$$g(r) = \det g_{\mu\nu} = F\det g_{mn}(r) = -FC^2(C+G)\;.$$

The tensor $g^{\mu\nu}$, which is inverse of the metric tensor ($g_{\mu\nu}g^{\nu\lambda} = \delta^\lambda_\mu$), has the form

$$g^{00} = \frac{1}{F(r)},\quad g^{0m} = 0,\quad g^{mn} = -\frac{1}{C(r)}\delta^{mn} + \frac{G(r)}{C(C+G)}\frac{x^m x^n}{r^2}\;. \quad (10)$$

The gravitational action (1) (with regard to the change of its sign) has the following form for the given metric

$$S_{gr} = \frac{c^4}{16\pi G}\int (R + Q)\sqrt{-g(r)}dt dx^1 dx^2 dx^3\;, \quad (11)$$

$$Q = \frac{\partial \Phi(r)}{\partial x^\mu} g^{\mu\nu} \frac{1}{2g(r)} \frac{\partial g(r)}{\partial x^\nu} = -\frac{\Phi'(r)g'(r)}{2(C(r)+G(r))g(r)} = -\frac{\Phi'(r)\Delta'(r)}{\Delta(r)^3} F(r)C(r)^2, \qquad (12)$$

where the notation for the volume factor is introduced

$$\Delta(r) = \sqrt{-g(r)} = C\sqrt{F(C+G)}$$

(the dash here and below in this section denotes differentiation by *r*).

The scalar curvature can be calculated using the known results of GR calculations [4, 5]. *In the original "Cartesian" coordinates (9), the scalar curvature $R(x^1, x^2, x^3)$ is not only generally covariant, but also form-invariant [5, ch. 13], so it is easier to calculate it in "spherical" coordinates.* The space-time interval (9) in "spherical" coordinates has the form [5, p. 176]

$$ds^2 = F(r)(dx^0)^2 - G(r)dr^2 - C(r)(dr^2 + r^2 d\theta^2 + r^2 \sin^2\theta \, d\varphi^2).$$

Let us redraft it as follows

$$ds^2 = F(r)(dx^0)^2 - A(r)dr^2 - r*^2(r)(d\theta^2 + \sin^2\theta \, d\varphi^2), \qquad (13)$$

with the below notations

$$A(r) = G(r) + C(r), \quad r^*(r) = rC^{1/2}(r).$$

*Unlike GR, the constraint restricts the group of admissible coordinate transformations and does not allow reducing the number of the sought metric components to two.*

The non-vanishing components of the connection, in the case of the metric (13), are somewhat different from the corresponding components of "standard" form [5, p. 178] (and coincide with them in the absence of a constraint, when $r^*(r) = r$)

$$\Gamma^t_{tr} = \Gamma^t_{rt} = \frac{F'}{2F}, \Gamma^r_{rr} = \frac{A'}{2A}, \Gamma^r_{\theta\theta} = -\frac{r*r*'}{A}, \Gamma^r_{\varphi\varphi} = -\frac{r*r*'\sin^2\theta}{A}, \Gamma^r_{tt} = \frac{F'}{2A},$$

$$\Gamma^\theta_{r\theta} = \Gamma^\theta_{\theta r} = \frac{r*'}{r*}, \Gamma^\theta_{\varphi\varphi} = -\sin\theta\cos\theta, \Gamma^\varphi_{\varphi r} = \Gamma^\varphi_{r\varphi} = \frac{r*'}{r*}, \Gamma^\varphi_{\varphi\theta} = \Gamma^\varphi_{\theta\varphi} = \text{ctg}\theta.$$

The expression for the curvature tensor changes accordingly. Using the expressions for the components of the connection, we find the scalar curvature in *"spherical" coordinates*

$$R(r,\theta,\varphi) = \frac{1}{2F}\left(\frac{F'}{A}\right)' + \frac{1}{2A}\left(\frac{F'}{F}\right)' + \frac{2}{r*^2}\left(\frac{r*r*'}{A}\right)' + \frac{2}{A}\left(\frac{r*'}{r*}\right)' - \frac{2}{r*^2} + \frac{2}{A}\left[\left(\frac{r*'}{r*}\right)^2 + \frac{r*'F'}{r*F}\right].$$

Singling out the terms that form a pure divergence, it is possible to write it in the form

$$R(r,\theta,\varphi) = \frac{1}{r*^2\sqrt{AF}}\frac{d}{dr}\left[r*^2\sqrt{AF}\left(\frac{F'}{AF} + \frac{4r*'}{r*A}\right)\right] - 2\left[\frac{r*'F'}{r*AF} + \frac{1}{A}\left(\frac{r*'}{r*}\right)^2 + \frac{1}{r*^2}\right]. \qquad (14)$$

Then it is necessary to note that, owing to the form-invariance of the scalar curvature, we see:

$$R(x^1, x^2, x^3) = R(r, \theta, \varphi).$$

In the presence of a constraint, it is more convenient not to proceed from the equations derived by varying the action by metric components, but to choose $\Delta(r)$ as one of the variable functions instead of $A(r)$.

Substituting the expression (14) for $R(x)$ and (12) for $Q(x)$ into the action, omitting the divergence and considering that $A = \Delta^2/C^2 F$, we derive

$$S_{gr} = \frac{c^4}{8\pi G}\int\left(\frac{\Delta}{r*^2} + \frac{r*^2 F}{\Delta r^4}(r*')^2 + \frac{1}{\Delta r^4}r*^3 r*' F' + \frac{\Phi'\Delta' r*^4 F}{2\Delta^2 r^4}\right)d^3 x dt.$$

Since all the functions in this expression depend on coordinates only in the combination $r(x)$, it is possible to pass to spherical coordinates in the volume integral, and the action will take the form

$$S_{gr} = \frac{c^4}{8\pi G}\int\left(\frac{\Delta}{r*^2} + \frac{r*^2 F}{\Delta r^4}(r*')^2 + \frac{1}{\Delta r^4}r*^3 r*' F' + \frac{\Phi'\Delta' r*^4 F}{2\Delta^2 r^4}\right)r^2\sin\theta \, dr d\theta d\varphi dt. \qquad (15)$$

One should note again that due to the restrictedly covariance of the scalar *Q*, *the whole action* is defined only for the metric (9). Moreover, the integration area is limited by an edge and is subject to further definition.

## 4. Derivation of gravity field equations

Let us introduce the variable $\xi = r^3$ instead of *r*, then the action (15) will take the form

$$S_{gr} = -\frac{3c^4}{8\pi G}\int\left(\frac{\Delta}{9r*^2} + \frac{Fr*^2}{\Delta}\left(\frac{dr*}{d\xi}\right)^2 + \frac{1}{\Delta}r*^3\frac{dr*}{d\xi}\frac{dF}{d\xi} + \frac{Fr*^4}{2\Delta^2}\frac{d\Phi}{d\xi}\frac{d\Delta}{d\xi}\right)d\xi\sin\theta \, d\theta d\phi dt.$$

The principle of least action will help one to come to gravity field equations

$$\frac{d}{d\xi}\left(\frac{r*^4 F}{\Delta^2}\frac{d\Delta}{d\xi}\right) = 0, \qquad (16)$$

$$-\frac{1}{9r*^2} + \frac{r*^2}{\Delta^2}\left(\frac{dr*}{d\xi}\right)^2 F + \frac{r*^3}{\Delta^2}\frac{dr*}{d\xi}\frac{dF}{d\xi} + \frac{1}{2\Delta^2}\frac{d}{d\xi}\left(r*^4 F \frac{d\Phi}{d\xi}\right) = 0, \qquad (17)$$

$$-\frac{r*^2}{\Delta}\left(\frac{dr*}{d\xi}\right)^2 + \frac{d}{d\xi}\left(\frac{r*^3}{\Delta}\frac{dr*}{d\xi}\right) - \frac{r*^4}{2\Delta^2}\frac{d\Delta}{d\xi}\frac{d\Phi}{d\xi} = 0, \qquad (18)$$

$$\frac{2\Delta}{9r*^3} + 2r*\frac{d}{d\xi}\left(\frac{r*F}{\Delta}\frac{dr*}{d\xi}\right) + r*^3\frac{d}{d\xi}\left(\frac{1}{\Delta}\frac{dF}{d\xi}\right) - 2\frac{r*^3 F}{\Delta^2}\frac{d\Delta}{d\xi}\frac{d\Phi}{d\xi} = 0. \qquad (19)$$

It follows from the equation (16)

$$\frac{r*^4 F}{\Delta^2}\frac{d\Delta}{d\xi} = \alpha, \qquad (16')$$

where $\alpha$ is a constant with length dimension.

## 5. Bringing the set of equations (17...19) to a self-conjugate form

Let us multiply the equation (17) by $2\Delta$, then subtract (18) multiplied by $2F$ from the result, and add the result to the equation (19) multiplied by $r*$; after simple transformations, the equation will be brought to a form:

$$\frac{d}{d\xi}\left[\frac{r*^4}{\Delta}\left(\frac{dF}{d\xi} + F\frac{d\Phi}{d\xi}\right)\right] = 0.$$

Hence

$$\frac{r*^4 F}{\Delta}\left(\frac{1}{F}\frac{dF}{d\xi} + \frac{d\Phi}{d\xi}\right) = \beta,$$

where $\beta$ is another constant with length dimension. Assuming $\beta = \sigma\alpha$ where $\sigma$ is a numerical factor and using (16'), this equation can be recorded as

$$\frac{1}{F}\frac{dF}{d\xi} + \frac{d\Phi}{d\xi} = \sigma\frac{1}{\Delta}\frac{d\Delta}{d\xi}, \quad \sigma = \frac{\beta}{\alpha}.$$

Considering that the function $\Phi(r)$ is determined correctly to a constant, we find

$$\Phi = -\ln(F\Delta^{-\sigma}). \qquad (17')$$

The equation (18) will be redrafted as follows

$$\frac{d}{d\xi}\left(\frac{r*^2}{\Delta}\frac{dr*}{d\xi}\right) = \frac{r*^3}{2\Delta^2}\frac{d\Delta}{d\xi}\frac{d\Phi}{d\xi}. \qquad (18')$$

After substituting this expression into the equation (19), it will take the form

$$r*^4\frac{d}{d\xi}\left(\frac{1}{\Delta}\frac{dF}{d\xi}\right) + 2r*^2\frac{d}{d\xi}\left(\frac{Fr*}{\Delta}\frac{dr*}{d\xi}\right) - 4\left[\frac{1}{\Delta}\left(\frac{r*dr*}{d\xi}\right)^2 + r*^2\frac{d}{d\xi}\left(\frac{r*}{\Delta}\frac{dr*}{d\xi}\right)\right]F + \frac{2\Delta}{9r*^2} = 0.$$

This equation is equivalent to the following

$$\frac{d}{d\xi}\left[\frac{r*^6}{\Delta}\frac{d}{d\xi}\left(\frac{F}{r*^2}\right)\right] + \frac{2\Delta}{9r*^2} = 0.$$

Integrating this equation by $\xi$, we derive

$$\frac{d}{d\xi}\left(\frac{F}{r*^2}\right) - \beta_1\frac{\Delta}{r*^6} + \frac{2\Delta}{9r*^6}\int_0^\xi \frac{\Delta}{r*^2}d\xi' = 0, \qquad (19')$$

where $\beta_1 = \left[\frac{r*^6}{\Delta}\frac{d}{d\xi}\left(\frac{F}{r*^2}\right)\right]_{\xi=0}$ is another constant with length dimension.

Thus, the general solution for the original set of equations depends on the choice of values of the three constants $\alpha$, $\beta_1$, $\sigma$.

## 6. Definition of integration constants

First of all, let us consider a case when $\alpha = 0$. In this case, it follows from (16')

$$\Delta(r) = const = \Delta(\infty) = 1.$$

Further, we find from (17'), (18'), (19')

$$\Phi(r) = -\ln F(r),$$

$$C^{\frac{1}{2}}(r) = \frac{r*(r)}{r} = const = \left.\frac{r*(r)}{r}\right|_{r=\infty} = 1,$$

$$F(r) = 1 - \frac{\beta_1}{r}, \quad A(r) = F^{-1}(r).$$

This solution aligns with the Schwarzchild solution when assigning the value $\beta_1$ to be equal to the gravitational radius of a body. However, in this case, the condition (5) is infringed since $\Delta(r) = 1$. Consequently, from the *point of view of a theory of gravity with a constraint,* this solution is deemed non-physical. Unlike GR, one should assume that $\beta_1 = 0$, resulting in the solution coinciding with the Minkowski metric.

Let us further assume that $\beta_1 = 0$, allowing the Minkowski metric to be a solution for this set of equations when the constant $\alpha$ is zero. If $\alpha$ is not equal to zero, then the set of equations will include a single-dimensional constant, which should be associated with the energy of the material body, creating non-homogeneity

$$\alpha = \frac{GE_{mat}}{c^4}. \tag{20}$$

This constant can be referred to as the *scale factor* of the gravity field.

Integrating the equation (19') once again, let us represent the function $F(r)$ in the form

$$F = \frac{2}{9} r*^2 \int_\xi^\infty \left( \int_0^{\xi'} \frac{\Delta}{r*^2} d\xi'' \right) \frac{\Delta}{r*^6} d\xi'. \tag{19''}$$

Substituting the expression for the derivative $\Delta(r*)$ in the equation (16'), we rewrite the equation (18') in the form

$$\frac{d}{dr*} \frac{1}{V} = \frac{3\alpha}{2r*F} \frac{d\Phi}{dr*}, \quad V = \frac{\Delta(r*)}{3r*^2} \frac{d\xi}{dr*}.$$

Proceeding from derivatives with respect to $\xi = r^3$ to derivatives with respect to $r*$ in all the ratios and introducing dimensionless coordinates $r/\alpha$ and $r*/\alpha$ (*retaining the former designations r and r* for them*), the initial set of equations can be recorded as follows:

$$\frac{1}{\Delta} \frac{d\Delta}{dr*} = \frac{3V(r*)}{Fr*^2}, \tag{21}$$

$$V(r*) = \frac{1}{1 - \frac{3}{2} \int_{r*}^\infty \frac{1}{r'*F} \frac{d\Phi}{dr'*} dr'*}, \quad \Phi = -\ln(F\Delta^{-\sigma}), \tag{22}$$

$$F(r*) = 2r*^2 \int_{r*}^\infty \left( \int_{r'*_{min}}^{r'*} V(r''*) \, dr''* \right) \frac{1}{r'*^4} V(r'*) \, dr'*, \tag{23}$$

$$\frac{\Delta(r*)r^2}{r*^2} \frac{dr}{dr*} = V(r*). \tag{24}$$

A non-zero value $r*_{min} = r*(0)$ generally means the existence of *an edge* of the space-time manifold.

Let us consider the metric behaviour at $r*_{min} = 0$ and small values of $r*$ (keeping in mind that these represent dimensionless values now). It follows from (23) that if there exists an integral

$$2 \int_0^\infty \left( \int_0^{r*} V(r'*) \, dr'* \right) \frac{V(r*)}{r*^4} dr* = b > 0,$$

then, having small $r*$, the function $F(r*) \approx b \times r*^2$. Then, assuming $V(r*) \approx b_1 \times r*^\nu \geq 0$, $\Delta(r*) \approx b_2 \times r*^\mu \geq 0$ and substituting these expressions into (21...23), we derive from the first equation

$$\nu = 3, \quad \mu = 3\frac{b_1}{b}.$$

Using L'Hopital's rules for evaluation of indeterminate forms, we derive the following from the second equation (22), when $r*$ tends to zero,

$$b_1 = \frac{2b}{2 - \sigma \times \mu}.$$

If we consider the last two ratios as equations with respect to $\mu$

$$\mu = \frac{6}{2 - \sigma \times \mu},$$

then there exists a bounded solution for any $\sigma \leq 1/6$

$$\mu = \frac{1 - \sqrt{1 - 6\sigma}}{\sigma} \leq 6.$$

Integrating the equation (24), having small values of $r, r*$, we come to

$$r^3(r*) = 3 \int_0^{r*} \frac{V(r'*)}{\Delta(r'*)} r'^{*2} \, dr'* \approx 3 \int_0^{r*} \frac{b_1}{b_2} r'^{*(5-\mu)} \, dr'* . \tag{25}$$

The last integral exists only at $\mu<6$. Let us now take into account that under these conditions

$$\Phi = -\ln(F\Delta^{-\sigma}) \approx -\ln(b \times b_2 \times r^{*(1+\sqrt{1-6\sigma})}) .$$

The scalar $\Phi$ tends to infinity when $r*$ tends to zero. It follows that the set of equations (21...24) can be resolved only when $r*_{min} = r*(0)$ is different from zero. This is possible only when the value of the constant included in the equation (22) is $\sigma =1/6$. Under this condition: $\mu=6$, $b_1=2b$, and (24) takes the form

$$r^3(r*) = 3 \int_{r*_{min}}^{r*} \frac{V(r*)}{\Delta(r*)} r*^2 \, dr* \approx 3 \int_{r*_{min}}^{r*} \frac{b_1}{b_2} \frac{dr*}{r*} \approx 3 \frac{b_1}{b_2} \ln \frac{r*}{r*_{min}} , \tag{26}$$

and with $r$ tending to zero, $r*$ tends to $r*_{min}$.

In the set of equations (21...24), the value $r*_{min}$ is an independent parameter, and additional considerations are necessary for its determination. Let us consider the ratio (7), coupling the energy of a non-homogeneous gravity field with the energy responsible for creating this non-homogeneity in a solid body. Substituting the expressions for the metric tensor components from (9) into (7), we derive

$$E_{mat} = \frac{c^4}{16\pi G} \int \sqrt{-g} \frac{1}{r^2} \frac{d}{dr} \left( \frac{r^2}{C+G} \frac{d\Phi}{dr} \right) dV = \frac{c^4}{4G} \left[ \int_0^\infty \sqrt{-g} \frac{d}{dr} \left( \frac{r^2}{(C+G)} \frac{d\Phi}{dr} \right) dr \right] . \tag{27}$$

Let us now take into account that by convention and also by virtue of the ratio (24)

$$C(r*) + G(r*) = \frac{r^4(\sqrt{-g})^2}{r*^4 F(r*)} .$$

Substituting this expression into (27), we derive:

$$E_{mat} = \frac{c^4}{4G} \left[ \frac{r*^2 F(r*)}{V(r*)} \frac{d\Phi}{dr*} \bigg|_{r*_{min}}^{r*\to\infty} - \int_{r*_{min}}^\infty \frac{r*^2 F(r*)}{V(r*)\sqrt{-g}} \frac{d\Phi}{dr*} \frac{d\sqrt{-g}}{dr*} dr* \right] .$$

By convention $\Delta(r*)$ and by virtue of the ratios (16'), (17'), we derive

$$E_{mat} = \frac{c^4}{4G} \left[ -\frac{r*^2}{V(r*)} \frac{dF}{dr*} \bigg|_{r*_{min}}^{r*\to\infty} - 3\alpha \times \ln F\Delta^{-\sigma}(r*_{min}) \right] . \tag{28}$$

In (28), the boundary values of the derivative function $F(r*)$ appear. In order for Newton's law to be effective for large distances where the gravity field is weak, the asymptotic behaviour of a metric should have the form [4, ch. XII]

$$g_{00} = F = 1 - \frac{2GE_{mat}}{c^4 r*} \ldots .$$

Considering this expression, we derive

$$\lim_{r*\to\infty} \left( -\frac{r*^2}{V(r*)} \frac{dF}{dr*} \right) = -\frac{2GE_{mat}}{c^4} ,$$

Having $r* = r*_{min}$ by virtue of the ratio (23)

$$\lim_{r*\to r*_{min}} \frac{dF}{dr*} = \frac{2F(r*_{min})}{r*_{min}} ,$$

and taking the above into account, we derive

$$-\frac{r*^2}{V(r*)} \frac{dF}{dr*} \bigg|_{r*_{min}}^{r*\to\infty} = -\frac{2GE_{mat}}{c^4} + \frac{2F(r*_{min})r*_{min}}{V(r*_{min})} .$$

Substituting this expression into (28) and proceeding to the dimensionless coordinate $r*/\alpha$ (*retaining its former notation $r*$*), we finally derive

$$2 = \frac{2F(r*_{min})r*_{min}}{3V(r*_{min})} - \ln F\Delta^{-\sigma}(r*_{min}) . \tag{29}$$

## 7. Approximate solution for the set of equations (21...24)

With the chosen initial value of $r*_{min}$, starting from the trial function $V^{(0)}(r*)$, we can derive the function $F^{(0)}(r*)$ from (23), and then derive $\Delta^{(0)}(r*)$ from (21) and $r^{(0)}(r*)$ from the equation (24). Consequently, we derive the new value $V^{(1)}(r*)$ from (22), etc.

The trial function, taking into account its behaviour at $r* \to \infty$, will be assigned as follows
$$V^{(0)}(r*) = 1 - \frac{v}{r*^2}, v = const, r* = [r*_{min}, \infty) . \tag{30}$$
Substituting this expression into (23), we derive
$$F^{(0)}(r*) = 1 - \frac{2}{3}\left(r*_{min} + \frac{v}{r*_{min}}\right)\frac{1}{r*} + \frac{2}{5}\left(r*_{min} + \frac{v}{r*_{min}}\right)\frac{v}{r*^3} - \frac{v^2}{3}\frac{1}{r*^4} . \tag{31}$$
For Newton's law to be effective at large distances, the following should be assumed
$$\frac{2}{3}\left(r*_{min} + \frac{v}{r*_{min}}\right) = 2 ,$$
hence
$$v = r*_{min}(3 - r*_{min}) .$$
Using this ratio, let us write the trial function in the form
$$V^{(0)}(r*) = 1 - \frac{r*_{min}(3 - r*_{min})}{r*^2}, r* = [r*_{min}, \infty) . \tag{32}$$
Accordingly
$$F^{(0)}(r*) = 1 - \frac{2}{r*} + \frac{6}{5}\frac{r*_{min}(3 - r*_{min})}{r*^3} - \frac{(r*_{min}(3 - r*_{min}))^2}{3}\frac{1}{r*^4} . \tag{33}$$
Substituting (30), (31) into (21), we derive
$$\Delta^{(0)}(r*) = \exp\left(-3\int_{r*}^{\infty}\frac{V^{(0)}(r'*)}{(r'*)^2 F^{(0)}(r'*)}dr'*\right) . \tag{34}$$
Substituting (32), (33), (34) into (29), we derive an equation to calculate $r*_{min}$. The solution of this equation is as follows
$$r*_{min}^{(0)} = 1.746 , \tag{35}$$
in this case
$$g_{00}(r*) = F^{(0)}(r*) = 1 - \frac{2}{r*} + \frac{6}{5}\frac{v}{r*^3} - \frac{v^2}{3}\frac{1}{r*^4}, v = 2.19 . \tag{36}$$
(It should be recalled here that, talking of the dimensionless value $r*$, we mean the dimensional value $r*$ divided by the scale factor $\alpha = \frac{GE_{mat}}{c^4}$.)

To complete the cycle, we calculate the following ratio based on the equation (24)
$$\frac{r^{(0)}(r*)}{r*} = \left(\frac{3}{r*^3}\int_{r*_{min}}^{r*}\frac{V^{(0)}(r'*)}{\Delta^{(0)}(r'*)}r'*^2 dr'*\right)^{1/3} . \tag{37}$$
To evaluate the error of the found analytical solution, let us compare it with the calculations of the following approximation. The dependences $V^{(0)}(r*)$ (line) and the next approximation $V^{(1)}(r*)$ (points), calculated on the basis of value $r*_{min}$ (35), are shown in Fig.1.
$$V^{(1)}(r*) = \left(1 + \frac{3}{2}\frac{1}{r*F^{(0)}(r*)} - \frac{3}{2}\int_{r*}^{\infty}\left(1 + \frac{V^{(0)}(r*)}{2r'*F^{(0)}(r'*)}\right)\frac{dr'*}{(r'*)^2 F^{(0)}(r'*)}\right)^{-1} .$$
Similar dependency pairs $F^{(0)}(r*)$, $F^{(1)}(r*)$, $\Delta^{(0)}(r*)$, $\Delta^{(1)}(r*)$ and $r_{rel}=r^{(0)}/r*$, $r^{(1)}/r*$ are given in Fig. 2, Fig. 3, Fig. 4, where
$$F^{(1)}(r*) = 2r*^2\int_{r*}^{\infty}\left(\int_{r'*_{min}}^{r'*}V^{(1)}(r''*)dr''*\right)\frac{1}{r'*^4}V^{(1)}(r'*)dr'* .$$
$$\Delta^{(1)}(r*) = \exp\left(-3\int_{r*}^{\infty}\frac{V^{(1)}(r'*)}{(r'*)^2 F^{(1)}(r'*)}dr'*\right) .$$
$$\frac{r^{(1)}(r*)}{r*} = \left(\frac{3}{r*^3}\int_{r*_{min}}^{r*}\frac{V^{(1)}(r'*)}{\Delta^{(1)}(r'*)}r'*^2 dr'*\right)^{1/3} .$$

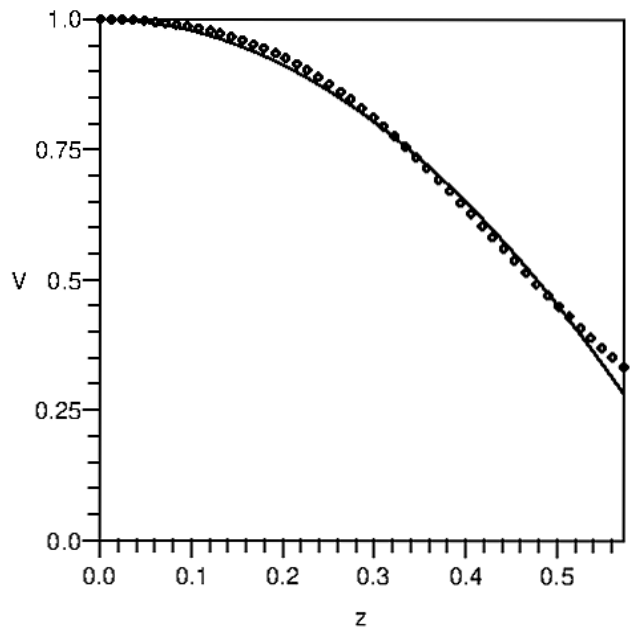

**FIG. 1.** The value $z = \alpha/r^*$ is plotted on the abscissa axis.

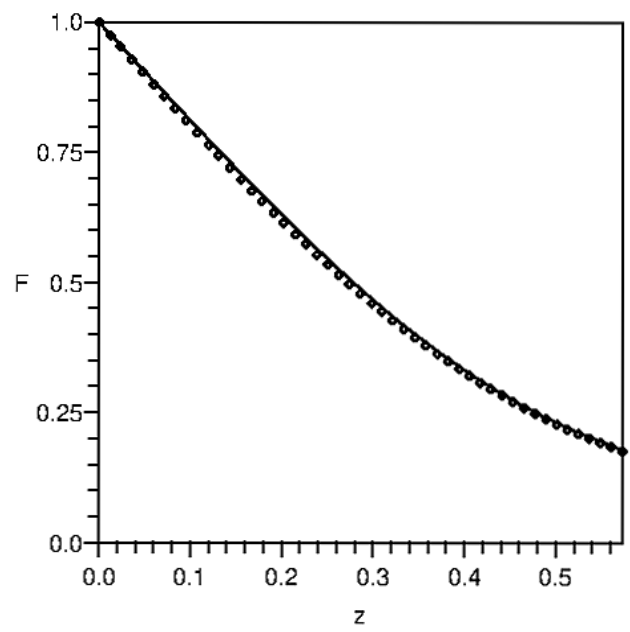

**FIG. 2.** The value $z = \alpha/r^*$ is plotted on the abscissa axis.

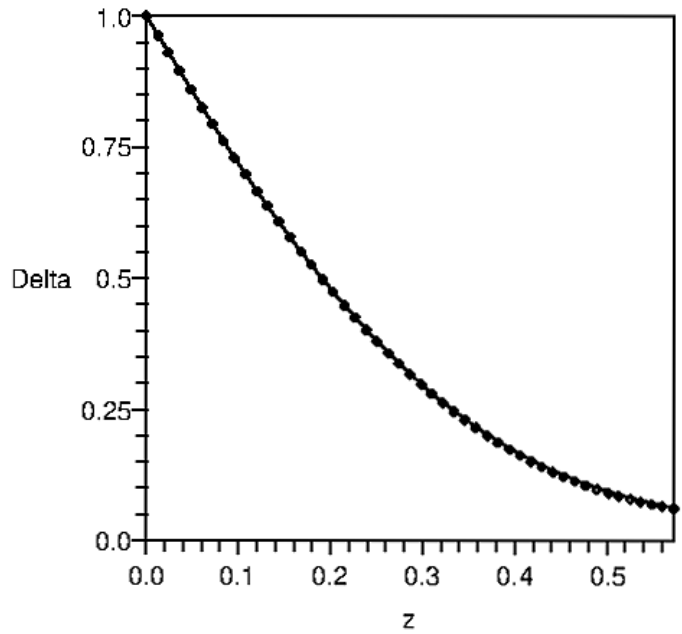

**FIG. 3.** The value $z = \alpha/r^*$ is plotted on the abscissa axis.

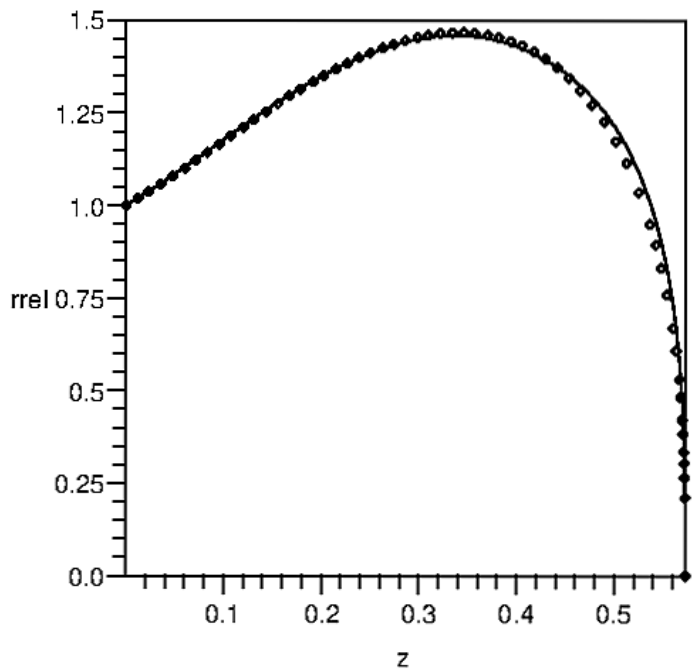

**FIG. 4.** The value $z = \alpha/r^*$ is plotted on the abscissa axis.

This is approximately the error in determining the higher terms of expansion, when describing the metric in the analytic form.

## 8. Discussion

In the theory of gravity with a constraint, only the covariance of gravity field equations is restricted, while the action and equations of matter fields remain generally covariant. This makes it possible to choose the most convenient form for the metric found, dictated by the problem to be solved. This can be shown by the example of the metric (9). After the solutions for the gravity field equations are found, we proceed to the spherical coordinates, and the space-time interval will take the form (14). Then we pass from coordinates ($r$, $\theta$, $\varphi$) to coordinates ($r^*(r)$, $\theta$, $\varphi$) using the ratio (24). After the appropriate transformation, we derive

$$ds^2 = F(r*)(cdt)^2 - \frac{V^2(r*)}{F(r*)} dr*^2 - r*^2 (d\theta^2 + \sin^2\theta d\varphi^2), r* = [\alpha r*_{\min}, \infty), \quad (38)$$

where the scale factor α is defined by the ratio (20), and the approximate function values are defined by the ratios (32...36). When r*>>α,

$$V(r^*) = 1,$$

with the accuracy to the terms of the second order of lowness, similarly

$$F(r^*) = 1 - 2\alpha/r^*,$$

with the accuracy of the terms of the third order of lowness. As a result, the expression for the interval (38) aligns with the solution for the GR equations found by Schwarzschild in 1916 for a metric outside the material body [4, ch. XII].

This implies that classical experiments verifying GR predictions, using calculations based on the Schwarzschild metric, will confirm the validity of both GR and the theory of gravity with a constraint. However, source [4, ch. XII]. indicates that all curvature tensor invariants for the Schwarzschild metric have a singularity at the point $r = 0$. Thus, this solution achieved within GR framework turns out to be non-physical.

The metric (38), unlike the "spherical" GR metric, never turns to zero. As can be seen from the presented drawings, it is continuous everywhere, has continuous derivatives up to the second order, and, consequently, finite curvature tensor invariants.

As it is widely known, the consequence of the GR's general covariance is the fact that unlike all other kinds of matter, solid, liquid, and gaseous media, the gravity field energy is not localisable (its energy-momentum density is equal to zero). The source [1] points out that this feature of GR violates the principle of material unity of the world. In this context, a question arises whether the theory of gravity constructed on the basis of the GR is a physical theory. The singularities inevitably arising in solving the equations of the theory of relativity [6] lead scholars to answer this question negatively. *The theory of gravity constructed on the basis of GR does not represent a physical theory.*

The gravity field in the theory of gravity with a constraint is endowed with all properties of the material medium. In this case, the number of equations exceeds the former by one, which makes the problem of solving gravitational equations more complicated in comparison with GR. Consequently, the general solution for this set of equations includes more integration constants. As demonstrated in [1] and in this paper, this enables the elimination of singularities during the solution process by appropriately choosing values for these integration constants.